\documentclass{elsart}
\usepackage{graphics}
\begin{document}
\hyphenation{com-bin-ed in-ter-ac-tio-on}
\begin{frontmatter}
\textbf{Preprint MPI~H-V5-2001}\\
\title{On the flavor mixing by a Light-Cone Hamiltonian 
       and the isotopic spin in QCD}
\author{Hans-Christian Pauli}
\address{Max-Planck Institut f\"ur Kernphysik, D-69029 Heidelberg, Germany}
\date{23 March 2001}
\begin{abstract}
    The structure of an effective light-cone Hamiltonian
    as recently derived is analyzed with emphasis on
    its prediction for flavor mixing in physical mesons.
    In a (perhaps over-simplified) model with one
    adjustable parameter, the empirical masses of
    all 25 pseudo-scalar mesons 
    which are possible for 5 flavors 
    are reproduced (almost) quantitatively.
    These results are coupled with explicit numerical
    estimates of flavor mixing. 
    In the present approach, the well-known mass degeneracy 
    of the pion triplet is caused by the mass degeneracy
    of the up and down quark.
\end{abstract}
\maketitle
\end{frontmatter}
%
\section{Introduction}
\label{sec:1}

In 1932, Heisenberg has postulated isotopic spin as a general symmetry
of nature to describe systematically atomic nuclei.
Isospin prevails to be important in hadronic physics,
both in concept \cite{DonGolHol92} and experiment \cite{Schroe00}.
But in the fundamental hadronic theory,
in the gauge theory of quantum chromodynamics (QCD),
isospin symmetry is not manifest in any obvious way.
The present work is intended to contribute yet an other facet
to an old and immanent question \cite{DonGolHol92}, 
without addressing however to be complete or exclusive.

Nowadays isospin symmetry is believed to be part of the more general 
chiral symmetry to the extent that the symmetry is thought to be 
manifest at sufficiently high temperatures and broken after a chiral 
phase transition in which the quarks acquire mass.
These ideas can be modeled by lattice gauge theory
(LGT) \cite{Schier00}. But when it comes down to calculate
hadrons in their ground state (at absolute zero),
LGT has some difficulties to calculate 
with the same quality of approximations for all the hadrons, 
because of the enormously different mass scales in the problem.
Particularly, it is not easy to extrapolate reliably down to the
very light and small mesons like the pions. 

Recently, an equally non-perturbative alternative 
has gradually emerged as reviewed in \cite{BroPauPin98}.
The light-cone approach addresses to diagonalize 
the (light-cone) Hamiltonian, 
$H_{LC}\vert\Psi\rangle = M^2\vert\Psi\rangle$,
and to calculate the spectra 
and invariant masses (squared) of physical particles.
In particular the method addresses to calculate the associated wave functions
$\Psi_{n}=\Psi_{q\bar q},\Psi_{q\bar q g},\dots$, 
which are the Fock-space projections of the hadrons eigenstate. 
The total wave function for a meson is then  
$ \vert\Psi_{meson}\rangle = \sum_{i} ( 
   \Psi_{q\bar q}(x_i,\vec k_{\!\perp_i},\lambda_i)  \vert q\bar q\rangle   +
   \Psi_{q\bar q g}(x_i,\vec k_{\!\perp_i},\lambda_i)\vert q\bar q g\rangle +
   \dots)$, for example. 

I present in section~\ref{sec:2} some general aspects of the method,
based on which I formulate in section~\ref{sec:2a} a sufficiently simple model. 
In order to be as concrete and pedagogic as possible, 
I continue in section~\ref{sec:3}
with an over-simplified model for 2 and 3 flavors which 
can be solved in closed form.   
I generalize in section~\ref{sec:4} to five flavors,
and compare to experiment.
In section~\ref{sec:5}, I draw the conclusions. 

\begin{figure} [t]
\begin{minipage}{65mm}
\begin{center}
  \resizebox{1.00\textwidth}{!}{%
  \includegraphics{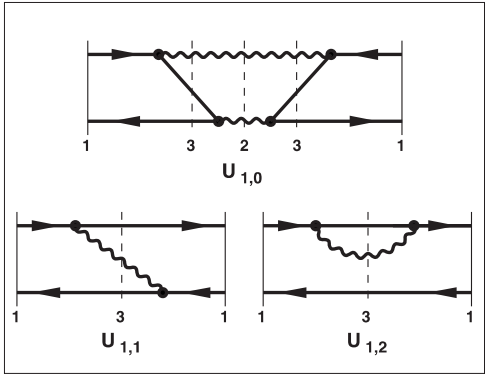} 
}\end{center} 
\caption{The three graphs of the effective interaction in the 
  $q\bar q$-space. 
  The lower two graphs correspond to the effective one-gluon-exchange
  interaction $U_\mathrm{ OGE}$,
  the upper corresponds to the effective two-gluon-annihilation 
  interaction $U_\mathrm{ TGA}$. 
  The figure is taken from Ref.\protect{\cite{Pau98}}.\label{fig:1}
}\end{minipage} 
\ { }\ \hfill
\begin{minipage}{65mm}
\begin{center}
\begin{tabular}{@{\hspace{.5ex}}c@{\hspace{.8ex}}|@{\hspace{.8ex}}
c@{\hspace{.8ex}}c@{\hspace{.8ex}}c@{\hspace{.8ex}}c@{\hspace{.8ex}}
c@{\hspace{.8ex}}c@{\hspace{.8ex}}c@{\hspace{.8ex}}c@{\hspace{.8ex}}
c@{\hspace{.1ex}}} 
     & $u\overline d$ & $u\overline s$ & $d\overline s$ 
     & $d\overline u$ & $s\overline u$ & $s\overline d$ 
     & $u\overline u$ & $d\overline d$ & $s\overline s$ \\ \hline
 $u\overline d$ &$E_1$ & 0    & 0    & 0    & 0    & 0    & 0    & 0    & 0   \\ 
 $u\overline s$ & 0    &$E_2$ & 0    & 0    & 0    & 0    & 0    & 0    & 0   \\
 $d\overline s$ & 0    & 0    &$E_3$ & 0    & 0    & 0    & 0    & 0    & 0   \\ 
 $d\overline u$ & 0    & 0    & 0    &$E'_1$& 0    & 0    & 0    & 0    & 0   \\ 
 $s\overline u$ & 0    & 0    & 0    & 0    &$E'_2$& 0    & 0    & 0    & 0   \\ 
 $s\overline d$ & 0    & 0    & 0    & 0    & 0    &$E'_3$& 0    & 0    & 0   \\ 
 $u\overline u$ & 0    & 0    & 0    & 0    & 0    & 0    &$e_4$ &$A_5$ &$A_6$\\ 
 $d\overline d$ & 0    & 0    & 0    & 0    & 0    & 0    &$A_5$ &$e_7$ &$A_8$\\ 
 $s\overline s$ & 0    & 0    & 0    & 0    & 0    & 0    &$A_6$ &$A_8$ &$e_9$\\ 
\end{tabular}
\end{center}
\caption{The kernel of $H_\mathrm{LC,eff}$ displayed as
   a block matrix illustrates the flavor mixing in QCD.
   $e_i\equiv E_i+A_i$. \label{fig:fockSp}
}\end{minipage} 
\end{figure}

\section{General considerations}
\label{sec:2}

Let me review in short some general aspects of the approach. 
The full light-cone Hamiltonian for gauge theory
with its complicated many-body aspects is reduced in \cite{Pau98}
by the method of iterated resolvents 
to the effective Hamiltonian  
\begin{equation}
   H_\mathrm{LC,eff} = T + U_\mathrm{OGE} + U_\mathrm{TGA}
.\label{eq:LCham}\end{equation}
By definition it acts only 
in the Fock space of a single quark and anti-quark. 
The kinetic energy $T$ be definition is diagonal and 
is the only part of the effective Hamiltonian
surviving in the limit of vanishing coupling constant.
Note that it has the dimension of a mass-squared, like
the other operators in the (light-cone) Hamiltonian.
The interaction (kernel) has three contributions,
which are displayed diagrammatically in Fig.~\ref{fig:1}. 

The diagrams in the figure are very compact. They 
are `energy' but not Feynman-diagrams;
all particle lines and all propagators are `effective' but well 
defined and represent summations over all orders \cite{Pau98}.
In diagram $U_{1,2}$, the effective gluon is absorbed
on the same line and does not change its kinematical state.  
It therefore generates effective masses
(in contrast to the bare Lagrangian ones). 
A certain part of them can be absorbed in $T$.
The one-gluon-exchange interaction $U_\mathrm{OGE}$ is
represented by diagram $U_{1,1}$.
Here, an effective gluon is emitted and absorbed on different lines which
causes a genuine interaction by the exchange of momentum.
The same holds true for the effective two-gluon annihilation 
interaction $U_\mathrm{TGA}$ correspoding to diagram $U_{1,0}$:
a $q\bar q$-pair of the same flavor is scattered into an other
pair with different momenta. The one-gluon-annihililation interaction
is absent in QCD, because a single gluon is colored and 
cannot couple to the color-neutral sector.
One should emphasize that the effective interaction
as obtained with the method of iterated resolvents \cite{Pau98}
has no operators which do not belong to one of the three classes 
of Eq.(\ref{eq:LCham}).

The structure of Eq.(\ref{eq:LCham}) has drastic consequences
whenever one considers a realistic case for more than 1 flavor.~--
Why is that?
Suppose, I was technically able to solve the effective Hamiltonian 
in Eq.(\ref{eq:LCham}) for just 1 flavor as a matrix diagonalization 
problem, as a warm-up exercise. 
Suppose, I want to treat next the case for 3 flavors. 
The matrix dimension increases by a factor $3\times 3\sim 10$, 
and the numerical effort for diagonalization on a computer 
increases thus by a factor 1000.
For the physical 6 flavors the effort is correspondingly larger.

But the symmetries in the effective Hamiltonian of Eq.(\ref{eq:LCham}) 
are quite helpful, as demonstrated in Fig.\ref{fig:fockSp}.
The matrix shown in this figure visualizes the kernel of the effective
Hamiltonian as a matrix of block matrices \cite{BroPauPin98}. 
Each block matrix represents a contribution from $H_\mathrm{LC,eff}$.
The symbol $(E_i)$ stands for contributions from $T + U_\mathrm{OGE}$. 
Most of the blocks are zero block-matrices, {\it i.e.}
all matrix elements inside a block vanish.
For example, the block $u\bar d \leftrightarrow u\bar s$ vanishes
because the Hamiltonian in Eq.(\ref{eq:LCham}) cannot connect them:
The kinetic energy cannot connect them since it is a diagonal operator;
the one-gluon-exchange interaction $U_\mathrm{OGE}$ cannot
connect them, since the $\bar d$ cannot change suddenly into an
$s$-anti-quark as seen from diagram $U_{1,1}$;
and, finally, the two-gluon-annihilation cannot connect them
since diagram $U_{1,0}$ requires the same flavor
on the left (and/or on the right).
The latter feature lets vanish also blocks like
$u\bar d \leftrightarrow s\bar s$.
This demonstrates that most of the Hamiltonian is reducible 
and that one can diagonalize blockwise.
Thus, only block matrix sectors like 
$u\bar u \leftrightarrow s\bar s$ are non-zero due to 
the diagrams $U_{1,0}$ and cause a mixing of flavors,
as consequence of QCD. 
They are denoted by $A_i$ in the figure.

It is thus reasonable to introduce a one-gluon-exchange
Hamiltonian and to diagonalize it on its own merit,
\begin{equation}
    H_\mathrm{OGE}    \,\vert\Psi_{f\bar f'}\rangle = 
    (T+U_\mathrm{OGE})\,\vert\Psi_{f\bar f'}\rangle =
    M_{f\bar f'}^2    \,\vert\Psi_{f\bar f'}\rangle 
,\label{eq:OGE-Ham}\end{equation}
to obtain flavor masses $M_{f\bar f'}$ and the associated wave functions
$\Psi_{f\bar f'}$.

\section{Formulation of the model}
\label{sec:2a}

Diagonalizing $H_\mathrm{OGE}$ and generating the many eigenfunctions
$\Psi_{f\bar f';i}$ can also be understood as the generation
of a unitary transformation to pre-diagonalize 
the flavor mixing matrix.
Although 
\begin{equation}
    \langle\Psi_{f\bar f;i}\vert U_\mathrm{TGA} 
    \vert\Psi_{f'\bar f';j}\rangle = 0
,\qquad \mbox{for }  i\neq j
,\label{eq:modI}\end{equation}
would be a false statement, in general,
one can expect that the off-diagonal matrix elements ($i-j$)
are (much) smaller than those on the diagonal ($i-i$).
Requiring Eq.(\ref{eq:modI}) to be true, however, 
makes things all of a sudden very simple:
The huge flavor-mixing matrix reduces to a state-by-state
diagonalization of a $n_f$ by $n_f$ flavor-mixing matrix $H_\mathrm{M}$,
where $n_f$ is the number of flavors.
Eq.(\ref{eq:modI}) will be refered to as model assumption I.

It is thus reasonable to introduce the ground-state-ground-state
correlations
\begin{equation}
    a_{ff'} \equiv
    \langle\Psi_{f\bar f}\vert U_\mathrm{TGA}\vert\Psi_{f'\bar f'}\rangle  
.\label{eq:5}\end{equation}
Since $m_u=m_d$, one has 
\begin{equation}
   a_{uu} = 
   a_{dd} = 
   a_{ud} = 
   a_{dd} \equiv a
,\qquad\mbox{and }
   M_{d\bar d} = M_{u\bar u}
.\end{equation}
I introduce as model assumption II 
\begin{equation}
   a_{us} = 
   a_{dc} = 
   a_{ub} = \dots = 
   a_{bb} \equiv a
,\label{eq:modII}\end{equation}
just to reduce their number.
In principle, the gs-gs correlations could be calculated from the 
wave functions, but below I will use $a$ as an adjustable parameter. 
It can be different for pseudo-scalar and vector mesons. 

\begin{table} [t]
\begin{minipage}[t]{68mm}
\caption{The calculated mass eigenvalues in MeV. 
   Those for singlet-1s states are given in the lower,
   those for singlet-2s states in the upper triangle.
   Taken from \protect{\cite{Pau00b}}. \label{tab:2.2}
}\begin{tabular}{c|rrrrrr} 
     & $\overline u$ & $\overline d$ 
     & $\overline s$ & $\overline c$ & $\overline b$ \\ \hline
    $u$ &         &      768&      871&     2030&     5418 \\
    $d$ &      140&         &      871&     2030&     5418 \\
    $s$ &      494&      494&         &     2124&     5510 \\
    $c$ &     1865&     1865&     1929&         &     6580 \\
    $b$ &     5279&     5279&     5338&     6114&          
\end{tabular}
\end{minipage} \hfill
\begin{minipage}[t]{68mm}
\caption{The empirical masses 
   of the flavor-off-diagonal physical mesons in MeV.
   The vector mesons are given in the upper, the scalar mesons
   in the lower triangle. \protect{\cite{Pau00b}}. \label{tab:1.2}}
\begin{tabular}{c|rrrrrr} 
     & $\overline u$ & $\overline d$ 
     & $\overline s$ & $\overline c$ & $\overline b$ \\ \hline
 $u$ &      & 768  & 892  & 2007 & 5325 \\ 
 $d$ & 140  &      & 896  & 2010 & 5325 \\ 
 $s$ & 494  & 498  &      & 2110 &  --- \\ 
 $c$ & 1865 & 1869 & 1969 &      &  --- \\ 
 $b$ & 5278 & 5279 & 5375 &  --- &      \\ 
\end{tabular}
\end{minipage} 
\end{table}

Solutions to Eq.(\ref{eq:OGE-Ham}) are actually available \cite{Pau00b},
within the so-called $\uparrow\downarrow$-model.
Its results are compiled in Table~\ref{tab:2.2}
and compared to the experimental masses in Table~\ref{tab:1.2},
taken from the particle data group \cite{PDG98}.
These data do not yet include the topped mesons,
which is the reason that the top quark is omitted here and below. 
The parameters of the model are the physical ones,
the strong coupling constant $\alpha$ and the quark masses $m_q$.
They are the same as in \cite{Pau00b} and 
tabulated in Table~\ref{tab:quarks}.
A similar model had been considered in earlier 
preliminary work \cite{Pau99e}.

For the present purpose, the codes for $\uparrow\downarrow$-model
have been run again to get the eigenvalues of Eq.(\ref{eq:OGE-Ham})
for the flavor diagonal case with no new parameters to adjust.
There are given in Tables~\ref{tab:4} and \ref{tab:5}.

\section{Flavor SU(2) and SU(3)}
\label{sec:3}

Let us restrict first to 2 flavors with equal masses 
$m_u=m_d$. 
The flavor-mixing matrix reduces to a 2 by 2 matrix, with
\begin{equation}
 H_\mathrm{M} = 
 \bordermatrix{%
         & u\bar u           & d\bar d           \cr 
 u\bar u & a + M^2_{u\bar u} & a \cr 
 d\bar d & a & a + M^2_{u\bar u} \cr 
}.\label{2eq:20}\end{equation}
The diagonalization of 
$H_\mathrm{M} \vert\Phi_{i}\rangle = M_{i}^2 \vert\Phi_{i}\rangle$ 
is easy. The two eigenstates, 
\begin{equation}
   \vert\Phi_{1}\rangle = \frac {1}{\sqrt{2}} 
   \pmatrix{\phantom{-}\vert u\bar u \rangle \cr
              -        \vert d\bar d \rangle \cr} 
,\ %
   \vert\Phi_{2}\rangle = \frac {1}{\sqrt{2}} 
   \pmatrix{\vert u\bar u \rangle \cr
            \vert d\bar d \rangle \cr} 
,\end{equation}
are associated with the eigenvalues
\begin{equation}
   M_{1}^2 = M^2_{u\bar u}  
,\qquad 
   M_{2}^2 = M^2_{u\bar u} + 2a  
.\end{equation}
The assumption of equal quark masses
leads thus to  $M_{u\bar d}=M_{d\bar u}=M_1$.
They can be arranged into a mass degenerate triplet of isospin 1,
independend of the numerical value of $a$.

\begin{table} [t]
\begin{minipage}[t]{32mm}
\caption{Model parameters: $\alpha=0.6904$, quark masses in MeV.
   \protect{\cite{Pau00b}}. \label{tab:quarks}
}\begin{tabular}{lr} 
 $q$ & $m_q$ \\
 $u$ &   406 \\
 $d$ &   406 \\
 $s$ &   508 \\
 $c$ &  1666 \\
 $b$ &  5054 \\
\end{tabular}
\end{minipage} \ \hfill
\begin{minipage}[t]{85mm}
\caption{The wave function of physical neutral pseudo-scalar mesons
   in terms of the $q\bar q$-wave functions.
   The leading component is normalized to $10$.\label{tab:wav}
}\begin{tabular}{lrrrrr} 
       { }      & $\pi^0$& $\eta$ & $\eta'$&$\eta_c$&$\eta_b$\\
 $u\overline u$ & 10.000 & -9.313 &  5.360 &  0.310 &  0.031 \\
 $d\overline d$ &-10.000 & -9.313 &  5.360 &  0.310 &  0.031 \\
 $s\overline s$ & -0.000 & 10.000 & 10.000 &  0.326 &  0.031 \\
 $c\overline c$ & -0.000 &  0.251 & -0.658 & 10.000 &  0.034 \\
 $b\overline b$ &  0.000 &  0.025 & -0.061 & -0.037 & 10.000 \\
\end{tabular}
\end{minipage} 
\end{table}
%

Next, consider 3 flavors. 
The flavor mixing matrix for the ground state becomes
\begin{equation}
 H_\mathrm{M} = 
 \bordermatrix{%
         & u\bar u & d\bar d & s\bar s \cr 
 u\bar u & a + M^2_{u\bar u} & a & a_{us} \cr 
 d\bar d & a & a + M^2_{u\bar u} & a_{ds} \cr 
 s\bar s & a_{us} & a_{ds} & a_{ss} + M^2_{s\bar s} \cr 
}.\end{equation}
The model assumption Eq.(\ref{eq:modII}) changes that into
a matrix with elements

\begin{equation}
 \langle f \vert H_\mathrm{M} \vert f'\rangle = 
 a + M_{f\bar f}^2\ \delta_{f,f'}
.\label{eq:20}\end{equation} 
If one assumes $m_u=m_d=m_c=m$, thus $M_{s\bar s}=M^2_{u\bar u}$,
as above, $H_\mathrm{M}$ can be diagonalized againin closed form. 
The three eigenstates  
\begin{equation}
   \vert\Phi_{1}\rangle =  \frac {1}{\sqrt{2}} 
   \pmatrix{\phantom{-}\vert u\bar u \rangle \cr
                     - \vert d\bar d \rangle \cr
                      0\vert s\bar s \rangle \cr } 
, 
   \vert\Phi_{2}\rangle = \frac {1}{\sqrt{6}} 
   \pmatrix{-\vert u\bar u \rangle \cr  
            -\vert d\bar d \rangle \cr
            2\vert s\bar s \rangle \cr} 
, 
   \vert\Phi_{3}\rangle =  \frac {1}{\sqrt{3}} 
   \pmatrix{\vert u\bar u \rangle \cr
            \vert d\bar d \rangle \cr
            \vert s\bar s \rangle \cr} 
,\label{eq:SU3}\end{equation}
are associated with the eigenvalues 
\begin{equation}
   M_{1}^2 = M^2_{u\bar u}  
,\qquad 
   M_{2}^2 = M^2_{u\bar u}  
,\qquad 
   M_{3}^2 = M^2_{u\bar u} + 3a  
.\end{equation}
The coherent state picks up all the strength, again. 
The eigenvalues of the remaining two states coincide with 
the unperturbated ones. 
State $\Phi_{1}$ can again be interpreted as 
the eigenstate for the charge neutral $\pi^0$ and the mass of
the coherent state $\Phi_{3}$ could be fitted with the $\eta'$.
But then state $\Phi_{2}$ is degenerate with the $\pi^0$:
Instead of a mass triplet one has a mass quadruplet.

Obviously, one cannot abstract from the appreciable mass difference
between the up and strange quark.
A $3\times 3$ matrix like in Eq.(\ref{eq:20}) can not be diagonalized 
in a simple way. 
It was diagonalized analytically but approximately in \cite{Pau99e}.

\section{Flavor SU(5), and its breaking by mass terms}
\label{sec:4}

%
\begin{table} [t]
\begin{minipage}[t]{65mm}
\caption{Compilation for the neutral pseudo-scalar mesons with  
   $a = (491\mbox{ MeV})^2$. Masses are given in MeV.
   \label{tab:4}
}\begin{tabular}{c|rrr} 
  \rule[-1em]{0mm}{1em}
  { }      & $M_{f\bar f}$ &    $M$ & $M_\mathrm{exp}$ \\ \hline
  \rule[1em]{0mm}{0.5em}
  $\pi^0$  &   140         &    140 &  135 \\ 
  $\eta $  &   140         &    485 &  549 \\ 
  $\eta'$  &   661         &$^*$958 &  958 \\ 
  $\eta_c$ &  2870         &   2915 & 2980 \\ 
  $\eta_b$ &  8922         &   8935 &  --- \\ 
\end{tabular}
\end{minipage}\ \hfill
\begin{minipage}[t]{65mm}
\caption{Compilation for the neutral pseudo-vector mesons with 
   $a = (255\mbox{ MeV})^2$. Masses are given in MeV.
   \label{tab:5} 
}\begin{tabular}{c|rrr} 
  \rule[-1em]{0mm}{1em}
  { }      & $M_{f\bar f}$ &    $M$ & $M_\mathrm{exp}$ \\ \hline
  \rule[1em]{0mm}{0.5em}
  $\rho^0$   &  768 &     768 &  768 \\ 
  $\omega$   &  768 &     832 &  782 \\ 
  $\Phi  $   &  973 &$^*$1019 & 1019 \\ 
  $J/\Psi $  & 3231 &    3242 & 3097 \\ 
  $\Upsilon$ & 9822 &    9825 & 9460 \\ 
\end{tabular} 
\end{minipage} 
\end{table}

Since one has to do numerical work anyway it is reasonable
to proceed immediately to five flavors.
The five flavor mixing-matrix $H_\mathrm{M}$
is given in Eq.(\ref{eq:20}), just with $n_f=5$.
The diagonal elements $M_{ff'}^2$ are obtained from Table~\ref{tab:4}.  
Diagonalizing $H_\mathrm{M} \vert\Phi\rangle = M^2 \vert\Phi\rangle$
numerically, produces the wavefunctions $\Phi$ in Table~\ref{tab:wav}
and the physical masses $M$ in Table~\ref{tab:4}.
The parameter $a$ is used to reproduce the mass of the $\eta'$,
as indicated by the star $^*$ in the table.
The corresponding results for pseudo-vector mesons are found 
in Table~\ref{tab:5}.

By adjusting one single parameter, one reproduces three empirical facts:
(1) the mass of the $\pi^0$ is (strictly) degenerate with $\pi^\pm$
(isospin);
(2) the unperturbed mass of the $\eta$ is lifted from the comparably 
small value of $140$~MeV to the comparatively large value of $485$~MeV;
(3) the unperturbed mass of the $\eta'$ is lifted by roughly
50\% the meet the experimental value.

The wave functions have also remarkable properties,
as seen from Table~\ref{tab:wav}.
The numerical results in the table remind to the SU(3) pattern
in Eq.(\ref{eq:SU3}). 
Particularly the isospin-pattern of the $\pi^0$ and the 
coherent-state pattern of the $\eta'$-wavefunction
should emphasized.
The heavy quark admixtures are small.

\section{Conclusion}
\label{sec:5}

In the present light-cone approach to gauge theory with an
effective interaction isospin is not a dynamic symmetry, 
but a consequence of equal up and down mass.
Flavor-SU(3) is an approximate symmetry.
The approach explains even the phenomenological observation that
flavor-SU(3) symmetry works better than SU(4) or SU(5);
the large mass of the heavy mesons dominates the flavor-mixing
matrix so strongly that the symmetry induced by the annihilation
interaction is destroyed. 
The present work contributes to the $\eta$-$\eta'$ 
puzzle \cite{BGPP97}
and exposes an accuracy comparable to state-of-art
lattice gauge calculations \cite{Kil97}.
To the best of my knowledge no other model including the
phenomenological ones \cite{DonGolHol92} 
covers the whole range of flavored hadrons with the same set of parameters.

The present approach is however in conflict with other theoretical constructs.
Zero modes are absent since one works with the light-cone gauge $A^+=0$
\cite{BroPauPin98}.
In consequence there are no chiral condensates which seem to be
so important otherwise. They are not needed here
since the parameter $a$ provides the additional mass scale.
It will be calculated from the theory in future work, 
removing then all parameter dependence beyond $\alpha$ and $m$.
At least one knows now that this is worth an effort.


\begin{thebibliography}{99}
\bibitem{DonGolHol92} 
     J.F. Donoghue, E. Golowich and B.R. Holstein,
     \textit{Dynamics of the standard model},
     Cambridge University Press, Cambridge, UK, 1992.
\bibitem{Schroe00}
     H. Schr\"oder,
     \textit{Experimental status of heavy mesons}
     Nucl. Phys. B (Proc. Supp.) \textbf{90} (2000) 74-82.
\bibitem{Schier00}
     G. Schierholz,
     \textit{Status and perspectives of Lattice QCD}
     Nucl. Phys. B (Proc. Supp.) \textbf{90} (2000) 207-213.
\bibitem{BroPauPin98} 
     S.J. Brodsky, H.C. Pauli, and S.S. Pinsky, 
     Phys. Rep. \textbf{301} (1998) 299-486. 
\bibitem{Pau98}
     H.C. Pauli,
     Eur. Phys. J. \textbf{C7} (1998) 289.
     hep-th/9809005.
\bibitem{Pau00b}
     H.C. Pauli,
     Nucl. Phys. B (Proc. Supp.) \textbf{90} (2000) 154-160.
     hep-ph/0103108. 
\bibitem{PDG98} 
     C. Caso {\it et al.} (Particle Data Group),
     Eur.Phys.J. \textbf{C3} (1998) 1. 
\bibitem{Pau99e}
     H.C. Pauli,
     Preprint MPI~H-V15-1999, May 1999, 4 pg. hep-ph/0103253.
\bibitem{BGPP97} 
     M. Burkardt, J. Goity, V. Papavassiliou, and S. Pate, Eds,
     \textit{The structure of the $\eta'$ meson},
     World Scientific, Singapore, 1997
\bibitem{Kil97} 
     G. Kilcup,
     p. 9-20,
     in: \textit{The structure of the $\eta'$ meson},
     M. Burkardt, J. Goity, V. Papavassiliou, and S. Pate, Eds,
     World Scientific, Singapore, 1997
\end{thebibliography}
\end{document}